# Laotian Bat Sarbecovirus BANAL-236 Uses ACE2 to Infect Cells by an Unknown Mechanism

## *BANAL-236 is the first infective coronavirus lacking eight obligate cis-acting UTR genomic secondary structures that are conserved in all known coronaviruses*

**By Steven C. Quay, MD, PhD[1]**
**Seattle, WA 98122 USA**

**Abstract.** A manuscript identified bat sarbecoviruses with high sequence homology to SARS-CoV-2 found in caves in Laos that can directly infect human cells via the human ACE2 receptor (Coronaviruses with a SARS-CoV-2-like receptor binding domain allowing ACE2-mediated entry into human cells isolated from bats of Indochinese peninsula, Temmam S., et al. https://www.researchsquare.com/article/rs-871965/v1). Here, I examine the genomic sequence of one of these viruses, BANAL-236, and show it has 5'-UTR and 3'-UTR secondary structures that are non-canonical and, in fact, have never been seen in an infective coronavirus. Specifically, the 5'-UTR has a 177 nt copy-back extension which forms an extended, highly stable duplex RNA structure. Because of this copy-back, the four obligate Stem Loops (SL) -1, -2, -3, and -4 *cis*-acting elements found in all currently known replicating coronaviruses are buried in the extended duplex. The 3'-UTR has a similar fold-back duplex of 144 nts and is missing the obligate poly-A tail. Taken together, these findings demonstrate BANAL-236 is missing eight obligate UTR *cis*-acting elements; each one of which has previously been lethal to replication when modified individually. Neither duplex copy-back has ever been observed in an infective sarbecovirus, although some of the features have been seen in defective interfering particles, which can be found in co-infections with non-defective, replicating viruses. They are also a common error seen during synthetic genome assembly in a laboratory. BANAL-236 must have evolved an entirely unique mechanism for replication, RNA translation, and RNA packaging never seen in a coronavirus and because it is a bat sarbecovirus closely related to SARS-CoV-2, it is imperative that we understand its unique mode of infectivity by a collaborative, international research effort.

**Introduction.** In September 2021, a non-peer reviewed preprint was published[2] that provided important evidence for the natural origin of SARS-CoV-2. The authors are associated with the Institut Pasteur, the century old Paris-based, infectious disease research institute where eight Nobel Prize winners did their groundbreaking work.

The authors reported finding five sarbecoviruses in caves in Laos, some of which have Spike Protein Receptor Binding Domains (RBD) which have high homology to SARS-CoV-2. The most remarkable finding reported in the pre-print was the apparent first ever observation of a bat sarbecovirus closely related to SARS-CoV-2 (BANAL-236) capable of directly infecting human cells via the ACE-2 receptor.

The finding of human ACE2 binding activity in bat viruses found in nature that was reported in this manuscript has been used to explain the finding that SARS-CoV-2 was remarkably well adapted for infectivity in humans from the very beginning of the pandemic.

The five BANAL strains described in the paper are:

---

[1] ORCID 0000-0002-0363-7651

[2] Coronaviruses with a SARS-CoV-2-like receptor binding domain allowing ACE2-mediated entry into human cells isolated from bats from the Indochinese peninsula

- Laotian *R. malayanus* BANAL-52, *R. pusillus* BANAL-103, and *R. marshalli* BANAL-236 coronaviruses, which are close to human SARS-CoV-2 and *R. affinis* RaTG13 coronaviruses phylogenetically.

- *R. malayanus* BANAL-116 and BANAL-247 coronaviruses, which belong to a sister clade with other bat coronaviruses (RmYN02, RacCS203, RpYN06, and PrC31) from different *Rhinolophus* species.

Here I examine the *cis*-acting genomic elements, sequences, and corresponding secondary structure of sarbecovirus BANAL-236. I find:

- BANAL-236 has a 177 nt 5'-UTR sequence, significantly longer than the typical 66-70 nt 5'-UTR found in sarbecoviruses. I am aware of no viable sarbecoviruses with a similar length 5'-UTR.
- BANAL-236 has inaccessible SL-1, -2, -3 (TRS-L), and -4 sequences buried in a 254 nt copy-back duplex 5' terminus. These conserved structures are found in all known sarbecoviruses and are required for infectivity.
- The BANAL-236 genome sequence also shows an unusual extension of the 3' terminus with another copy-back sequence, this one of 114 nts.
- The BANAL-116 genome sequence is lacking a TRS-L sequence as it seems to be missing an approximately 250 nt 5' sequence, when aligned to the other BANAL virus sequences. The 5' genome sequence of coronaviruses taken from bat fecal samples are often difficult to obtain and its absence in BANAL-116 is noted here not as an anomaly but for completeness of this report.

It has been over 40 years since research on the *cis*-acting elements of coronavirus replication began.[3] BANAL-236 is the first coronavirus to have evolved an unknown, independent mechanism of infectivity.

**Methods.** This analysis is based on the GenBank sequence for BANAL-236 and BANAL-116. Currently there are no SRA files available for either GenBank accession, making further analysis impossible. Because of the unusual findings here, I wrote an email to the corresponding author to have him confirm that BANAL-236 could infect VERO cells and that the reported sequence in GenBank was correct, which he did.[4] Sequence alignment was performed using BLAST and SnapGene software (Version 6.0). Secondary structures were determined using the RNAfold Web Server.

**Results.**

**BANAL-236 has an elongated 5'-UTR**

In sarbecoviruses, the length of the initial 5'-UTR from the capped end to the Leader-Transcription Regulatory Sequence (TRS-L; the universally conserved 7 nt sequence, AACGAAC) has been reported to be approximately 70 nt.[5]

[3] Schochetman G, Stevens RH, Simpson RW (1977) Presence of infectious polyadenylated RNA in the coronavirus avian bronchitis virus. Virology 77:772–782.

[4] Personal communication with M. Eliot.

[5] Nagy PD, Simon AE. 1997. New insights into the mechanisms of RNA recombination. Virology 235:1–9; Dufour D, Mateos-Gomez PA, Enjuanes L, Gallego J, Sola I. 2011. Structure and functional relevance of a transcription-regulating sequence involved in coronavirus discontinuous RNA synthesis. J. Virol. 85:4963–73.

To confirm this, the initial 5'-UTR for the 44 curated sarbecoviruses, selected by Jungreis et al.,[6] that are at evolutionary distances well-suited for identifying protein-coding genes and non-coding purifying selection were examined, as shown in the Text-Figure below. The data also included several viruses with 5'-UTR lengths ranging from 7 to 55 which are not shown in the Figure.

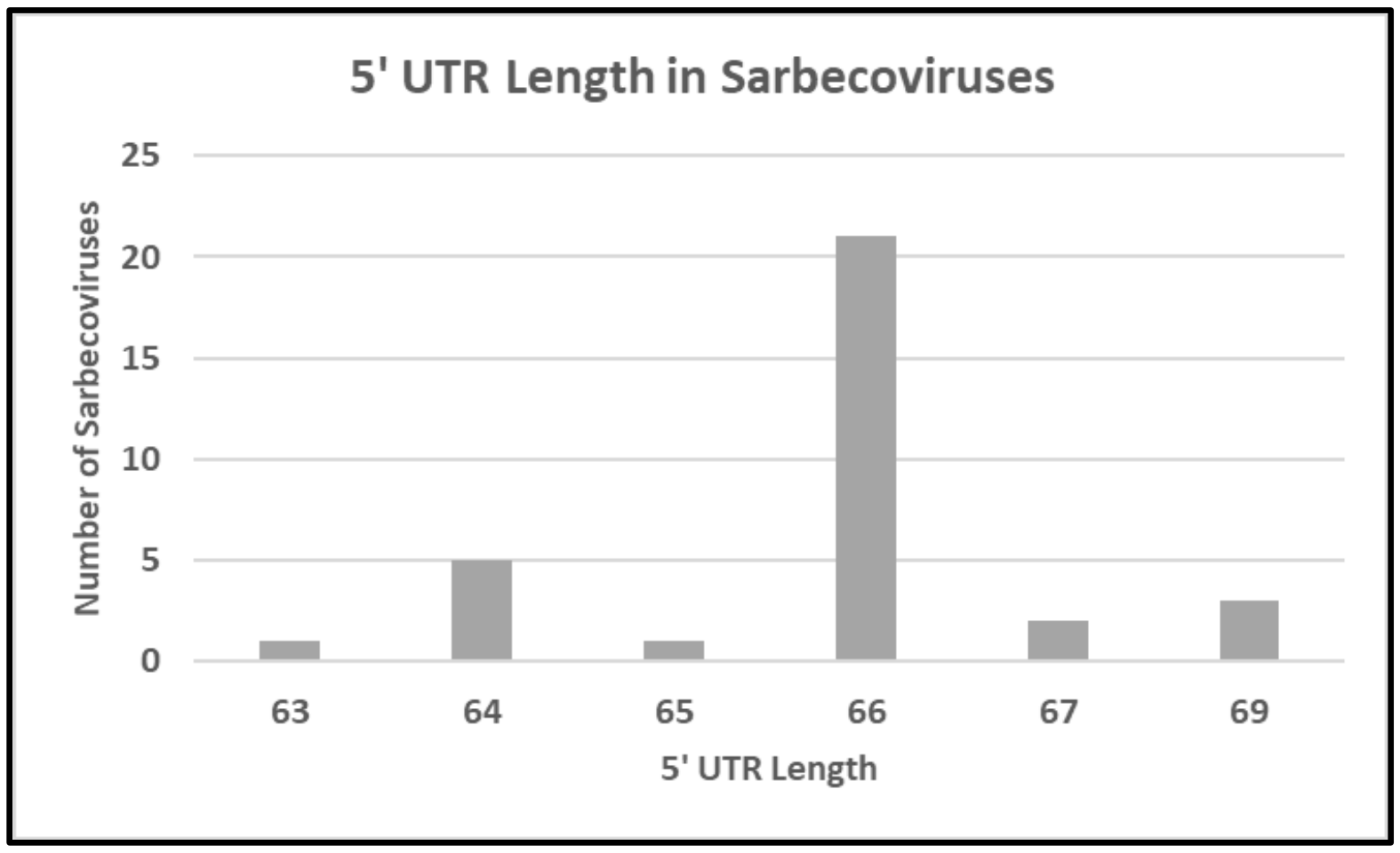

It is likely the few foreshortened 5'UTRs were simply an artefact of sample collection in the wild; that is, a consequence of the known difficulty of sequencing the 5' end of viruses collected from fecal samples.

To examine the 5'-UTR for BANAL-236, an alignment of all five BANAL viruses was performed in SnapGene (below). As seen, BANAL-236 has a 133 nt extension not seen in the other BANAL viruses and has a total length from the 5'-terminus to the TRS-L (red bracket below) of 177 nts.

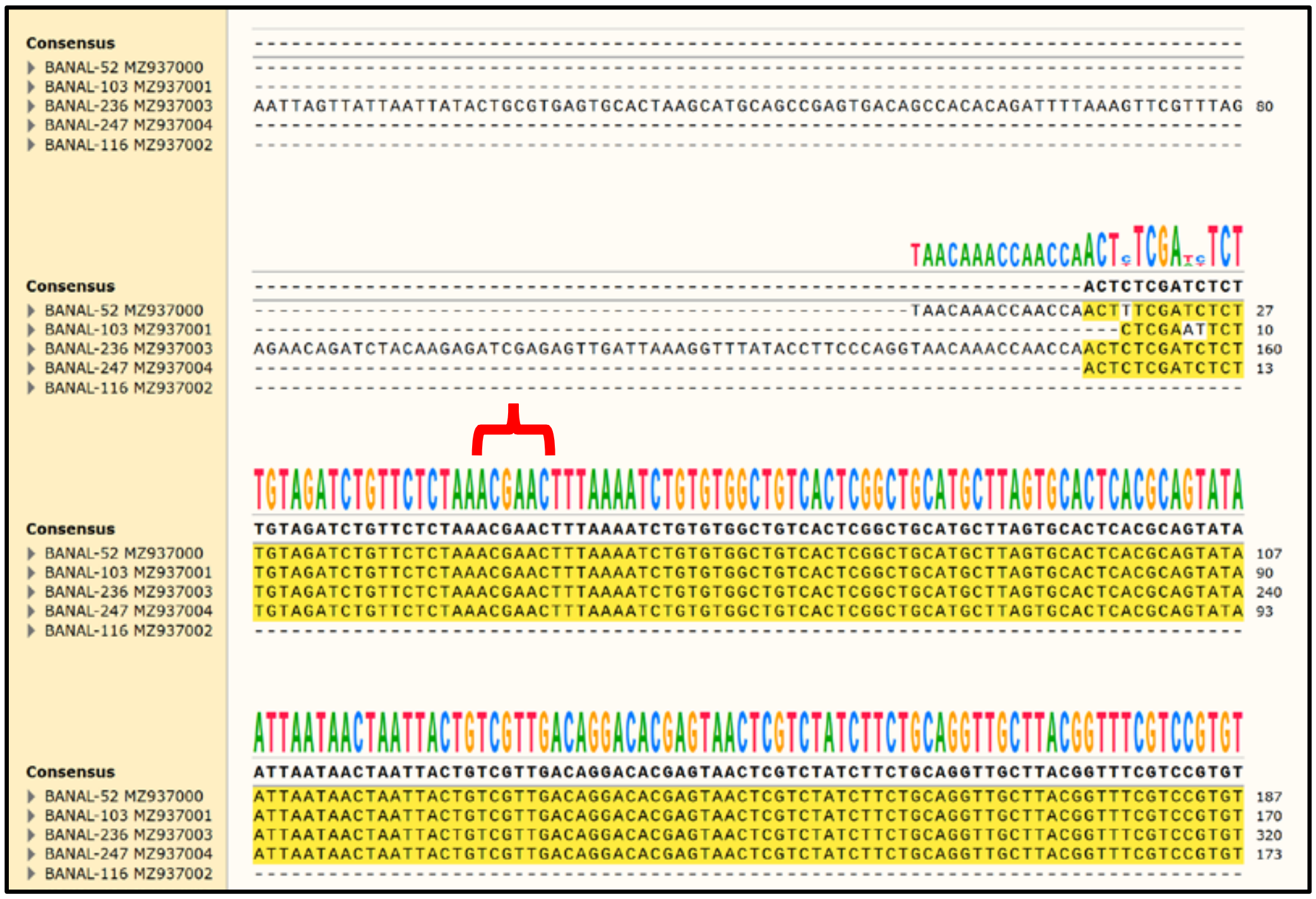

6 Jungreis, I., Sealfon, R. & Kellis, M. SARS-CoV-2 gene content and COVID-19 mutation impact by comparing 44 Sarbecovirus genomes. Nat Commun 12, 2642 (2021). https://doi.org/10.1038/s41467-021-22905-7

To determine the likelihood of finding a sarbecovirus with a 177 nt 5’-UTR in nature, the UTR lengths from the Jungreis et al. paper were first transformed to account for the skewness in the data (noted above). The skewness was treated by seeking a data transformation that achieved approximate normality using the Box-Cox family of power transformations. This turned out to be $X^{6.2097}$. With this transformation it is possible to evaluate the likelihood of finding a natural virus with a 5’-UTR length of 177. Using this transformation, a z-value of 1427 is calculated.

This translates into a probability of essentially zero for the likelihood of finding this structure in a natural sarbecovirus. Obviously, the known under sampling of sarbecoviruses from nature renders this probability determination less meaningful.

**The BANAL-236 5’-UTR is an extended copy-back duplex sequence never seen before**

A BLAST analysis of the 133 nt extension shows, of course, the 133 nt identity of the BANAL-236 virus but surprisingly also shows a plus/minus, “copy-back,” match within the early 5’ sequences of BANAL-236 (bottom alignment in Figure below).

```
Bat coronavirus isolate BANAL-20-236/Laos/2020, complete genome
Sequence ID: MZ937003.1  Length: 30024  Number of Matches: 2

Range 1: 1 to 133  GenBank  Graphics                    ▼ Next Match  ▲ Previous Matc

Score           Expect   Identities        Gaps          Strand
246 bits(133)   3e-61    133/133(100%)     0/133(0%)     Plus/Plus

Query  1    AATTAGTTATTAATTATACTGCGTGAGTGCACTAAGCATGCAGCCGAGTGACAGCCACAC  60
Sbjct  1    ............................................................  60

Query  61   AGATTTTAAAGTTCGTTTAGAGAACAGATCTACAAGAGATCGAGAGTTGATTAAAGGTTT  120
Sbjct  61   ............................................................  120

Query  121  ATACCTTCCCAGG  133
Sbjct  121  .............  133


Range 2: 143 to 254  GenBank  Graphics      ▼ Next Match  ▲ Previous Match  ▲ First Matc

Score           Expect   Identities        Gaps          Strand
202 bits(109)   7e-48    111/112(99%)      0/112(0%)     Plus/Minus

Query  1    AATTAGTTATTAATTATACTGCGTGAGTGCACTAAGCATGCAGCCGAGTGACAGCCACAC  60
Sbjct  254  ............................................................  195

Query  61   AGATTTTAAAGTTCGTTTAGAGAACAGATCTACAAGAGATCGAGAGTTGATT  112
Sbjct  194  .................................................G..  143
```

**The 5’UTR of BANAL-236 is a 112 nt reverse complement of critical portions of the *cis*-acting elements found downstream**

An alignment was performed with SARS-CoV-2 and the four closest bat coronaviruses with high homology to SARS-CoV-2: BANAL-52, RaTG13, BANAL-103, and BANAL-236. As can be seen in the Text-Figure below, the 112 nt sequence found at the beginning of BANAL-236 (blue arrow) ends at the beginning of the traditional 5’-UTR of SARS-CoV-2 and RaTG13. From there, while BANAL-236 has high sequence homology to SARS-CoV-2, beginning at BANAL-236 nt 146 and continuing to nt 254 (red arrow) there is a 100% homologous reverse complement 109 nt segment to the initial segment that is within critical *cis*-acting elements. As will be seen below, these two complementary reverse sequences form a highly stable 112 nt copy-back structure, destroying the normal, conserved secondary structure found in all viable coronaviruses.

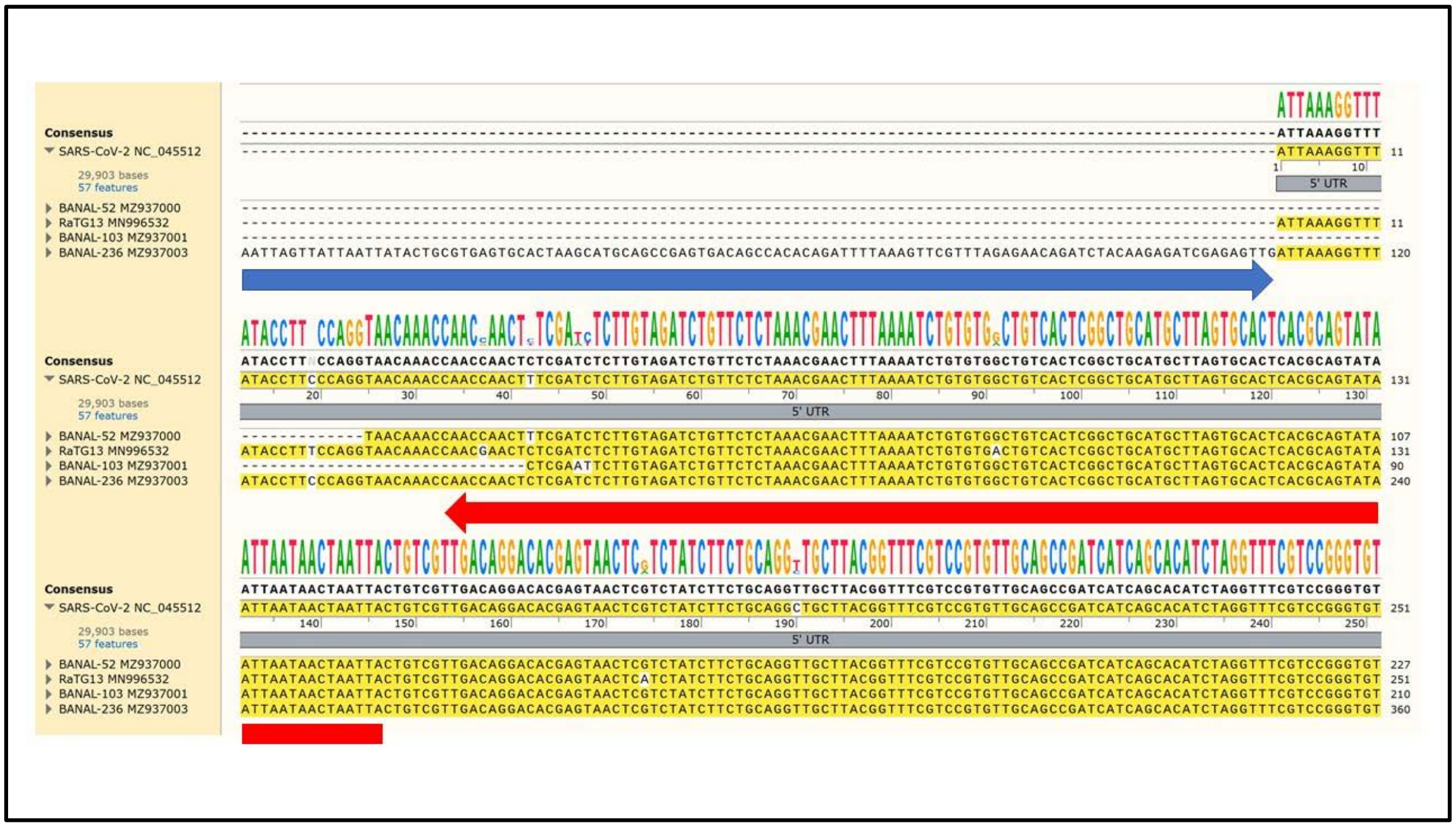

**The 5'- and 3'-UTR sequences of sarbecoviruses, including SARS-CoV-2, includes important conserved, obligate secondary structures which form *cis*-acting elements necessary for viral infectivity**

The 5'- and 3'-UTR of sarbecoviruses play important roles in infectivity, including being involved with replication, translation, and viral packaging of the positive-strand RNA during viroid production.[7] An analysis of the secondary structure of the 301 nt sequence in the beginning of the SARS-CoV-2 genome has been made, using inline and RNase V1 probing.[8] The resulting structure is shown in this Text-Figure from Reference 8. As can be seen there is a complex set of secondary structures, with their role in viral infectivity noted. These structures, but not necessarily sequences, are conserved in all coronaviruses; for example, elements without sequence homology can be exchanged across genera and can produce infective viruses.

[7] Miao Z, Tidu A, Eriani G, Martin F. Secondary structure of the SARS-CoV-2 5'-UTR. *RNA Biol*. 2021;18(4):447-456. doi:10.1080/15476286.2020.1814556; Yang D, Leibowitz JL. The structure and functions of coronavirus genomic 3' and 5' ends. *Virus Res*. 2015;206:120-133. doi:10.1016/j.virusres.2015.02.025

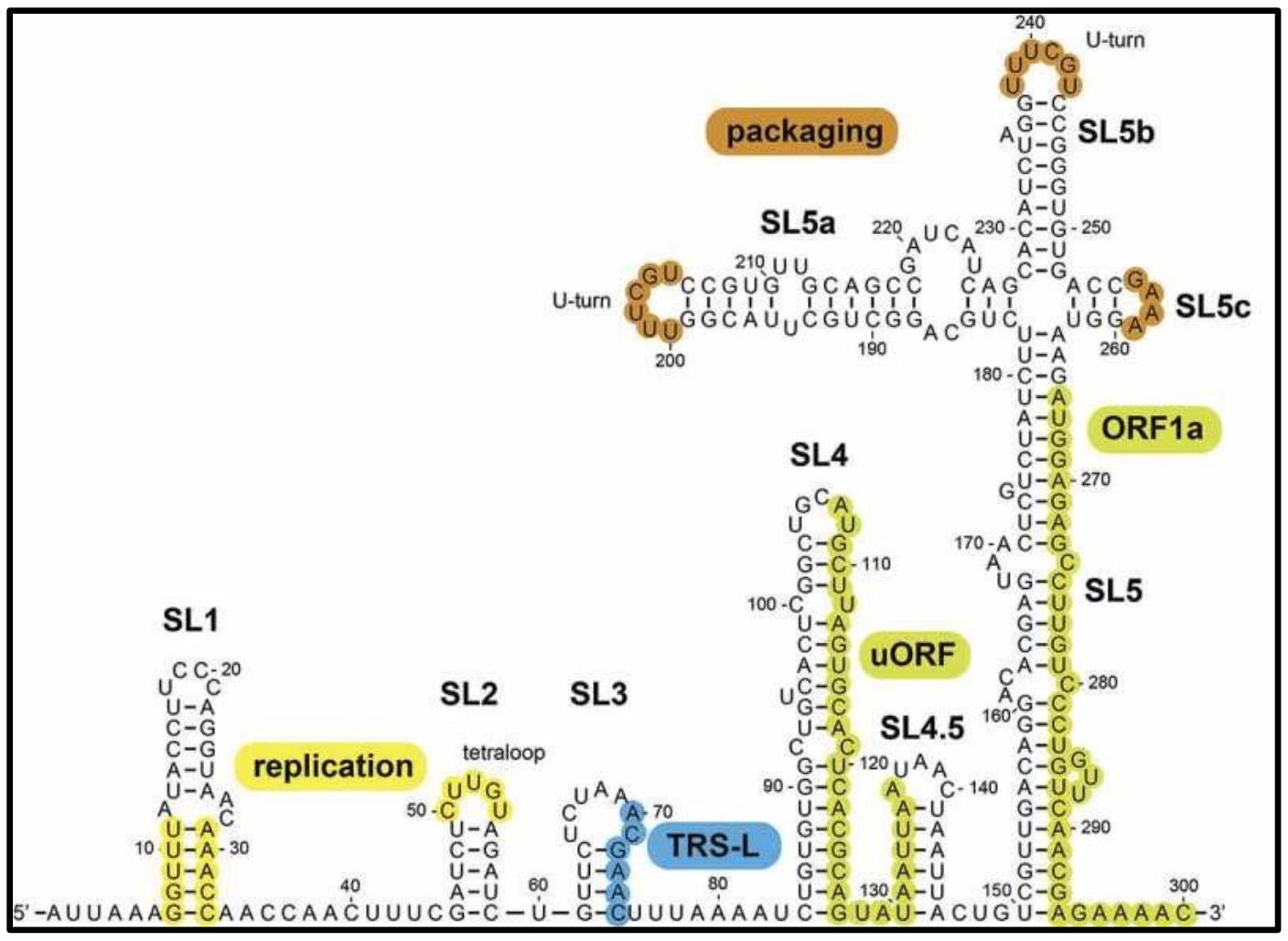

To validate a non-experimental, computer-based RNA secondary structure algorithm,[9] the same 301 nt region of SARS-CoV-2 was input to the algorithm and the structure shown below was generated.

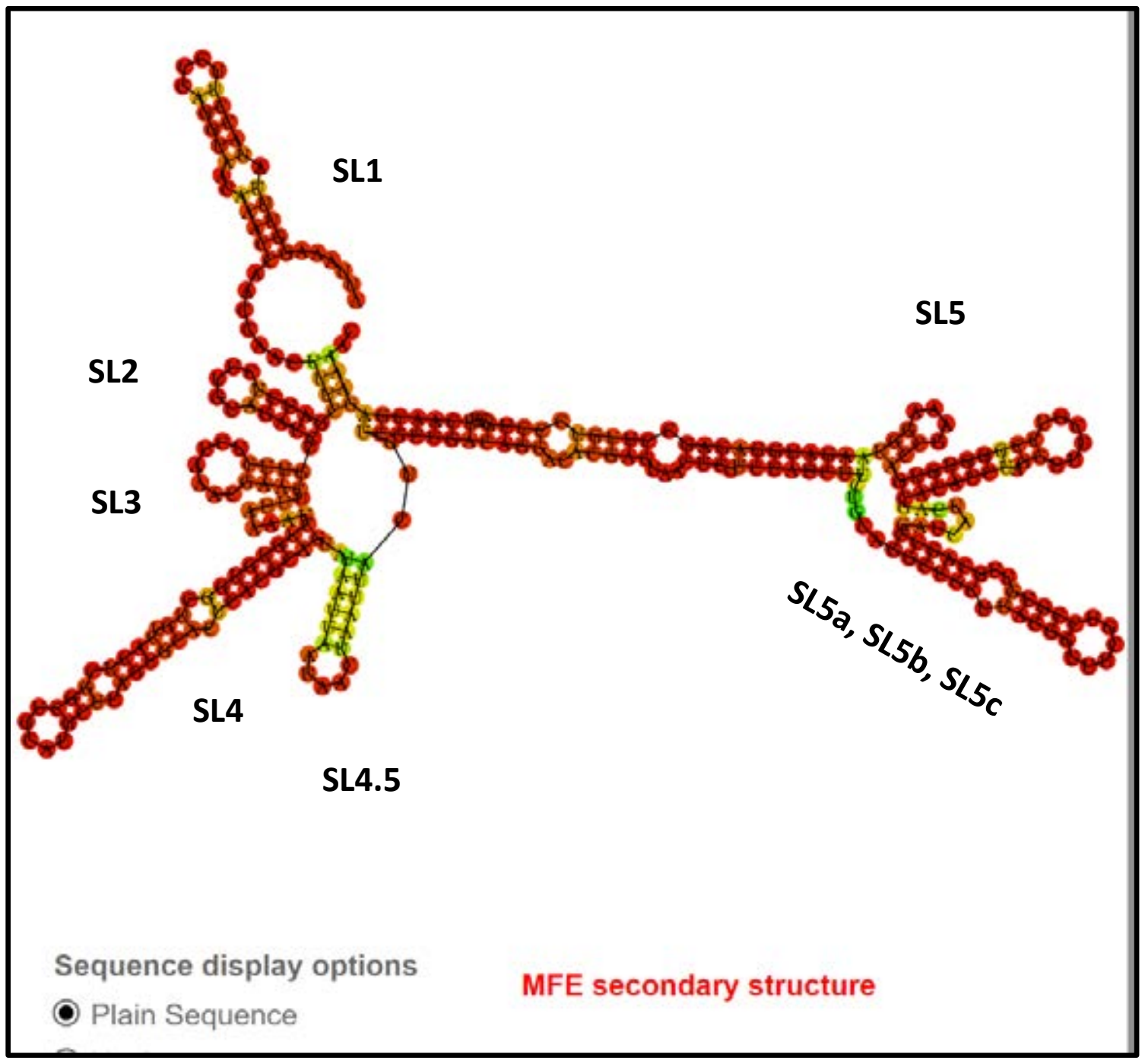

As can be seen above, the five SL structures identified by the computer-based algorithm correspond exactly with the experimentally determined structure, indicating the computer algorithm is capable of generating real world secondary structures.

**Natural leader transcriptional regulatory sequences (TRS-L) have an open, hairpin secondary structure critical for subgenomic RNA synthesis**

[9] RNAFold Web Server ; Huston NC, Wan H, de Cesaris Araujo Tavares R, Wilen C, Pyle AM. Comprehensive in-vivo secondary structure of the SARS-CoV-2 genome reveals novel regulatory motifs and mechanisms. Preprint. *bioRxiv*. 2020;2020.07.10.197079. Published 2020 Jul 10. doi:10.1101/2020.07.10.197079

The transcription process in beta coronaviruses is controlled by transcription-regulating sequences (TRSs) located about 70 nt from the 5' end of the genome, the leader sequence (TRS-L) and identical sequences preceding each viral gene (TRS-B). TRSs are a conserved core sequence and in sarbecoviruses it is always the seven nt sequence, AACGAAC. Transcription requires sequence identity between donor and acceptor RNAs, and thus the hairpin structures present in the acceptor RNA provide the necessary non-base paired sequence.[10]

Secondary structure analysis of the TRS-L region[11] shows that the TRS-L is exposed in a **loop, structured hairpin** that is relevant for replication and transcription. The figure below shows the required hairpin loop of the TRS-L, with a required 3 nt base-paired stem and demonstrates the mechanism of mRNA synthesis. This is the SL3 of the above Figures.

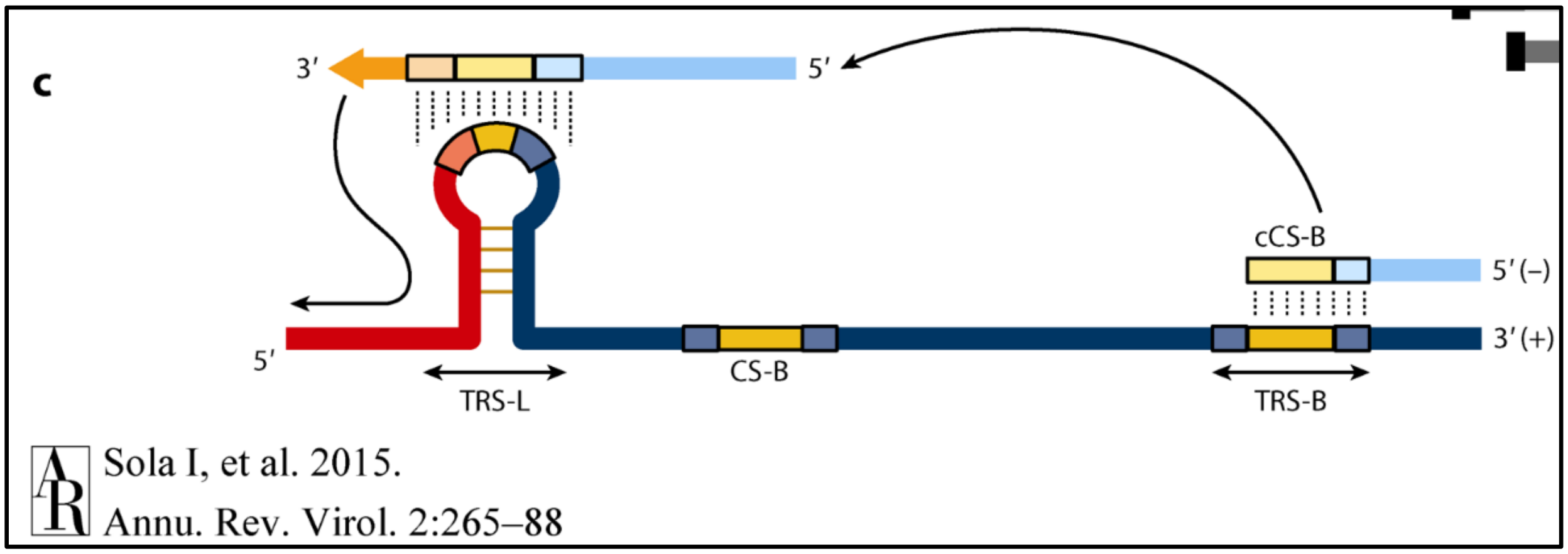

**BANAL-236 has a non-canonical 5'-UTR secondary structure**

The secondary structure of the 301 nts of the 5' sequence of BANAL-236 is shown below and contains none of the SL1 to SL5 structures, including an absence of the TRS-L structure.

[10] Nagy PD, Simon AE. 1997. New insights into the mechanisms of RNA recombination. Virology 235:1–9.
[11] Dufour D, Mateos-Gomez PA, Enjuanes L, Gallego J, Sola I. 2011. Structure and functional relevance of a transcription-regulating sequence involved in coronavirus discontinuous RNA synthesis. J. Virol. 85:4963–73.

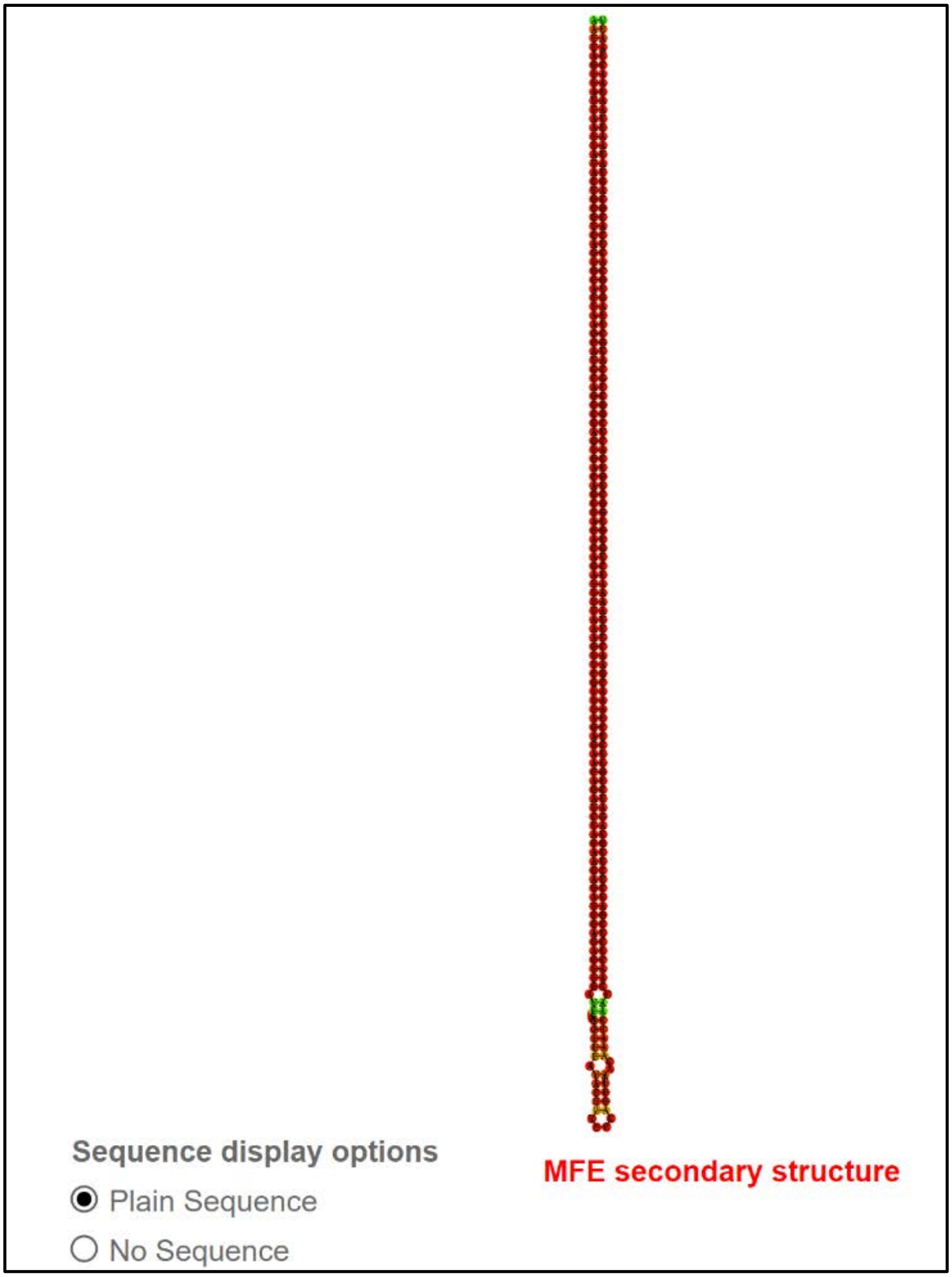

The closest structure to the TRS-L of BANAL-236 that has been studied is the modification of the 3 nt stem duplex to a 9 nt stem duplex in the transmissible gastroenteritis coronavirus.[12] While this still contained the hairpin structure allowing TRS-B and TRS-L to begin to seed the annealing of the nascent RNA molecules, the additional stability of the stem caused by changing from a 3 nt duplex to a 9 nt duplex lead to a reduction in subgenomic synthesis to 0.3% of the wildtype level.

[12] David Dufour, Pedro A. Mateos-Gomez, Luis Enjuanes, José Gallego, and Isabel Sola. Structure and Functional Relevance of a Transcription-Regulating Sequence Involved in Coronavirus Discontinuous RNA Synthesis. J of Virology, Vol 85, 15 May 2011, P 4963-4973. https://doi.org/10.1128/JVI.02317-10; Alexander O. Pasternak, Erwin van den Born, Willy J. M. Spaan, and Eric J. Snijder. The Stability of the Duplex between Sense and Antisense Transcription-Regulating Sequences Is a Crucial Factor in Arterivirus Subgenomic mRNA Synthesis. J of Virology, Vol 77, 15 January 2003, P. 1175-1183. https://doi.org/10.1128/JVI.77.2.1175-1183.2003; Alonso S, Izeta A, Sola I,

There is no known mechanism for a coronavirus with the 5'-UTR secondary structure of BANAL-236 to be an infective virion, yet its infection of VERO cells has been verified.

**The 3'-UTR contains a similar, unnatural duplex secondary structure never seen before**

The 3' terminus of the five BANAL viruses is shown in the Textbox below.

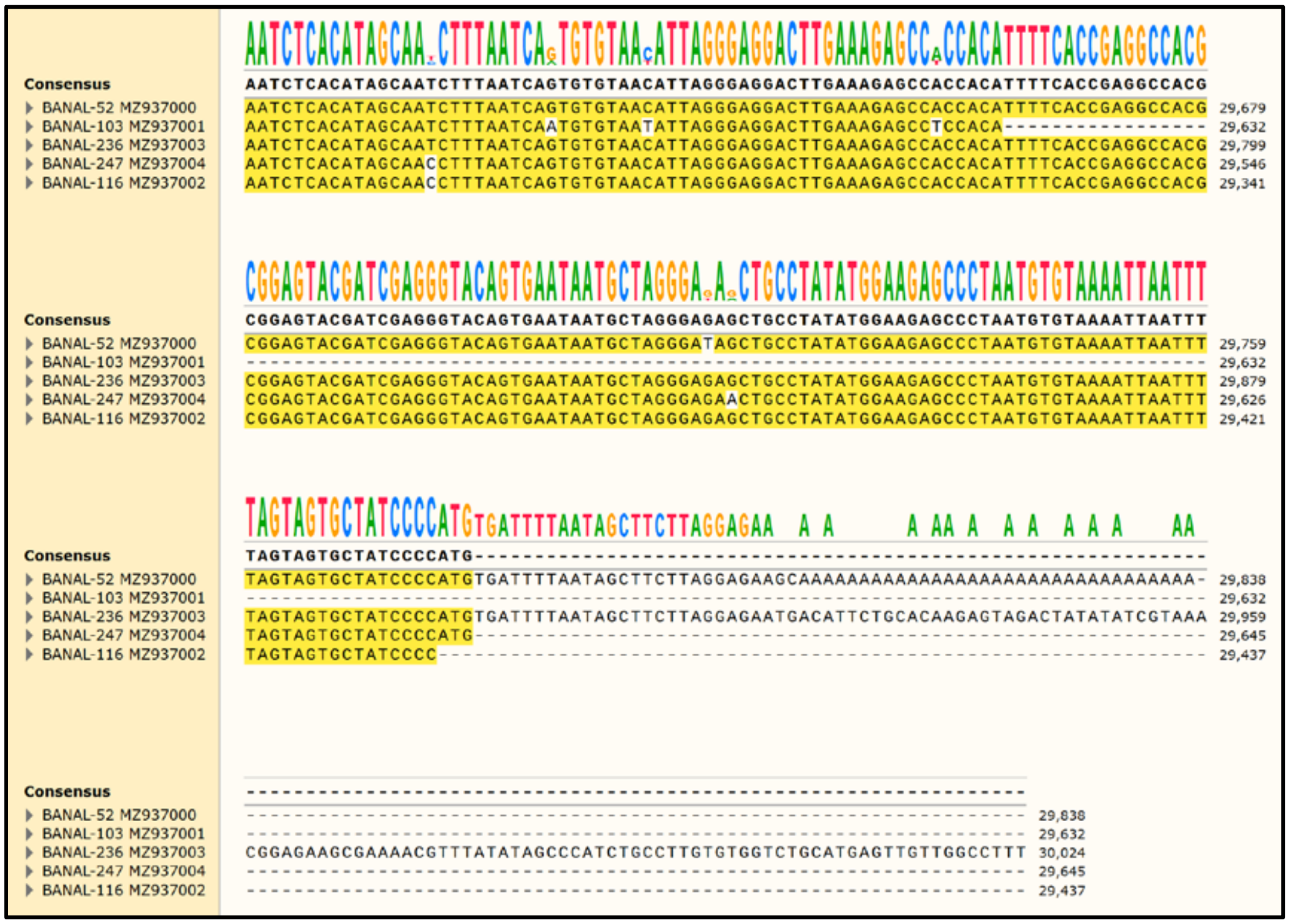

A SnapGene analysis (below) of BANAL-236, starting with the N gene and extending to the 3' terminus (positions 28,331 to 30,022), shows similar findings to the 5'-UTR sequence just described. Beginning within the end of the N gene and extending for 102 nts is a segment (blue arrow) that has a reverse complement at the end of the genome (red arrow). This will form a stable copy-back duplex that disrupts the 3'-UTR *cis*-acting elements. Of note, the obligatory poly-A genome tail is also missing.

Enjuanes L. Transcription regulatory sequences and mRNA expression levels in the coronavirus transmissible gastroenteritis virus. *J Virol*. 2002;76(3):1293-1308. doi:10.1128/jvi.76.3.1293-1308.2002; Yiyan Yang, Wei Yan, A Brantley Hall, Xiaofang Jiang, Characterizing Transcriptional Regulatory Sequences in Coronaviruses and Their Role in Recombination, *Molecular Biology and Evolution*, Volume 38, Issue 4, April 2021, Pages 1241–1248, https://doi.org/10.1093/molbev/msaa281

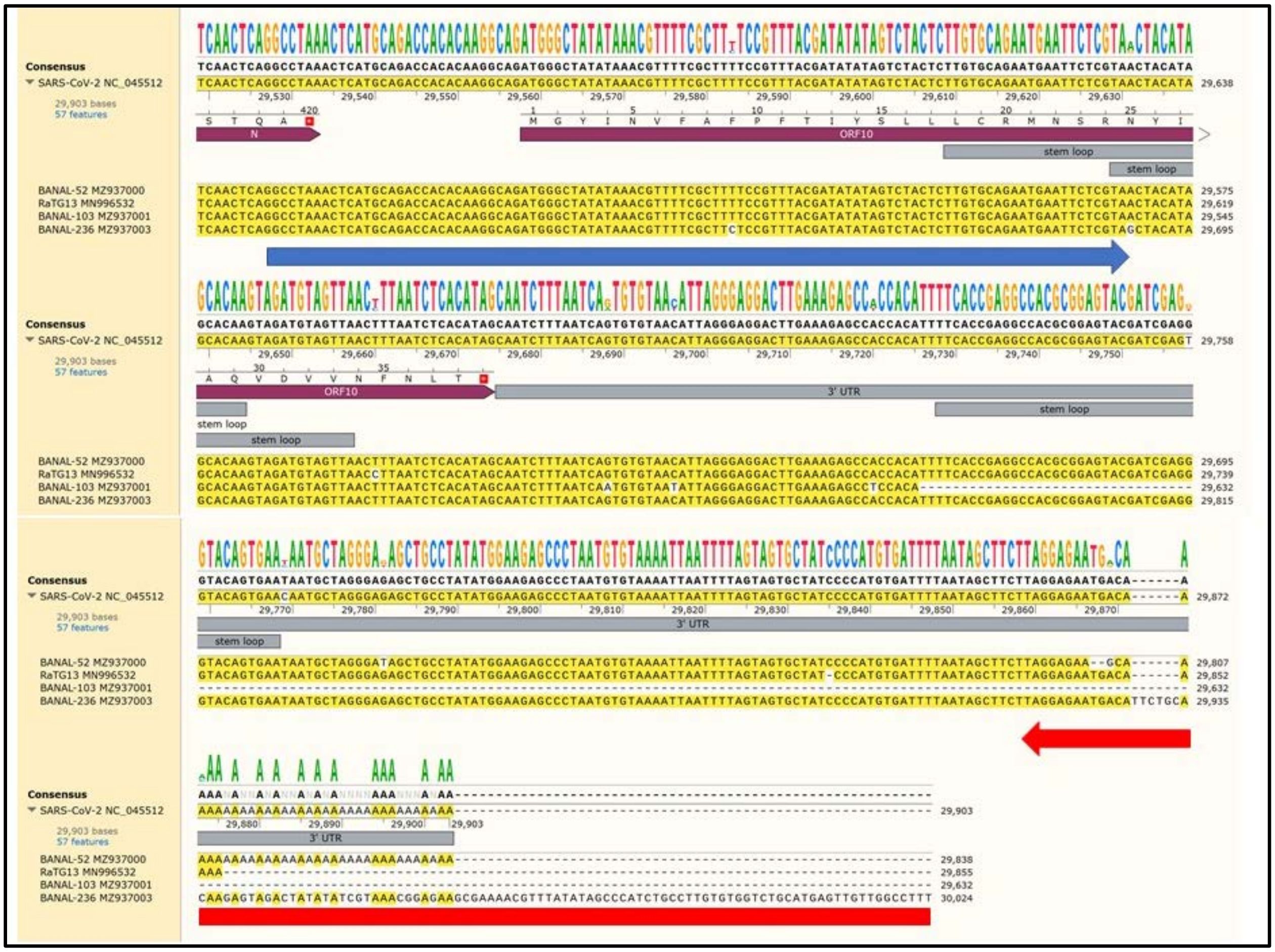

The secondary structure of the 3' sequence 29,583-30,024 (441 nts) of BANAL-236 (left, below) is compared to the 441 nt terminus of SARS-CoV-2 (right, below).

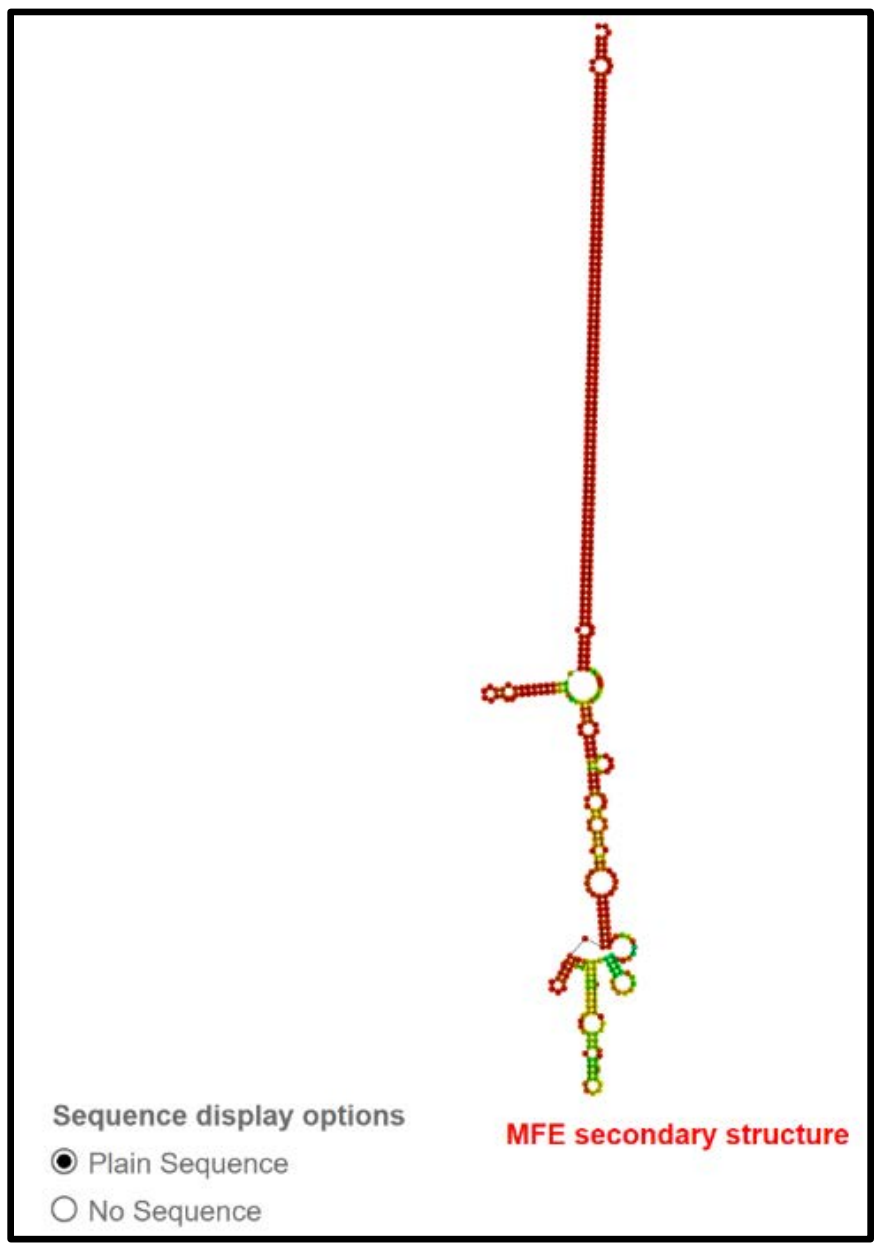

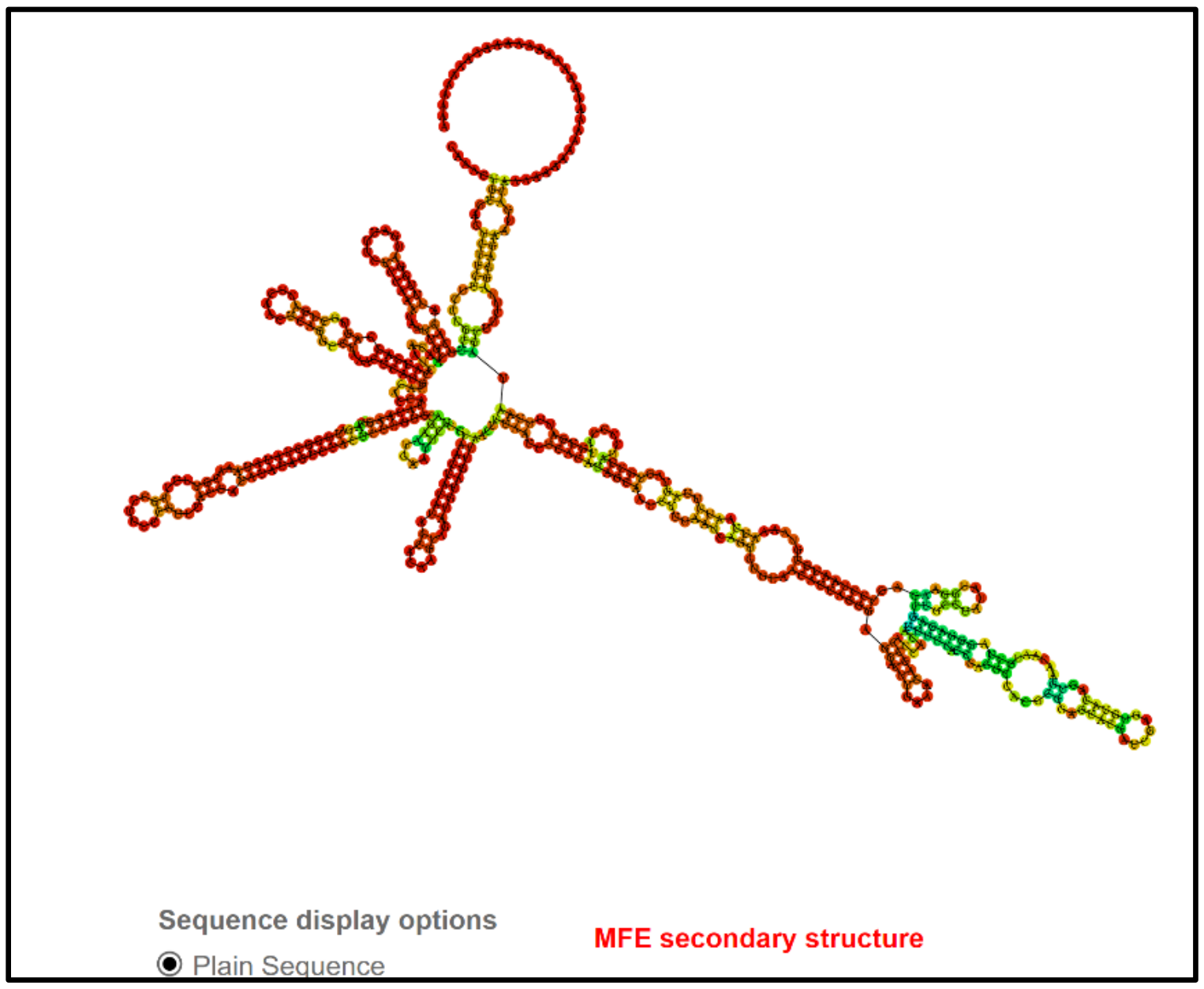

As can be seen, only SARS-CoV-2 has the complex secondary structure previously seen in infective coronaviruses.

## Discussion

### BANAL-236 is missing eight UTR *cis*-elements previously believed essential for coronavirus infectivity

Coronavirus research over the last forty years has led to a consensus of the structure and role of highly conserved *cis*-acting UTR secondary structures in replication, RNA translation, subgenomic mRNA synthesis, and viral RNA packaging.[11] Numerous experiments have demonstrated that viral particles missing even one of these essential elements lack infectivity.

Because BANAL-236 is the first bat virus with high homology to SARS-CoV-2 that can directly infect human cells via the ACE2 receptor special attention has been directed to it. Here I show that it has both 5'- and 3'-UTR copy-back structures which form extended, structureless duplex RNA in locations where complex secondary structures were expected to be seen and which are necessary for successful infectivity.

The missing 5'- and 3'-UTR secondary structures have previously been shown to be required for infectivity. Examining the details of secondary structure, BANAL-236 is missing eight secondary structures required for coronavirus viability (Text-Table below). In each case, the deletion of even one of these elements individually has been lethal to viral replication.

Here eight separate elements are simultaneously deleted and yet BANAL-236 has been demonstrated to be infective.

| UTR | Name | Location in SARS-CoV-2 | Function | Reference |
|---|---|---|---|---|
| 5'-UTR | Stem Loop 1 (SL1) | 7-33 | A bipartite domain architecture for SL1, which was previously reported to play a role in coronavirus replication | Structural lability in stem-loop 1 drives a 5' UTR-3' UTR interaction in coronavirus replication. Li L, Kang H, Liu P, Makkinje N, Williamson ST, Leibowitz JL, Giedroc DP J Mol Biol. 2008 Mar 28; 377(3):790-803. |
| | Stem Loop 2 (SL2) | 45-59 | A pentaloop is stacked atop a 5-bp stem; important for sgRNA synthesis; highly conserved in all beta coronaviruses | The solution structure of coronaviral stem-loop 2 (SL2) reveals a canonical CUYG tetraloop fold. Lee CW, Li L, Giedroc DP FEBS Lett. 2011 Apr 6; 585(7):1049-53. |
| | Transcription Regulatory sequence (SL3) | 61-75 | The transcription regulatory sequence (TRS) is a conserved feature of β-coronaviruses and is required for sgRNA production. | The structure and functions of coronavirus genomic 3' and 5' ends. Yang D, Leibowitz JL Virus Res. 2015 Aug 3; 206():120-33. |
| | Stem Loop 4 (SL4) | 84-128 | The SL4 of SARS-CoV-2 adopts a bipartite domain structure. Deletion of MHV 5'UTR SL4 is lethal and the genome carrying this deletion is defective in directing subgenomic RNA synthesis. A viable mutant in which SL4 was replaced with a sequence unrelated stem-loop supports the hypothesis that SL4 functions in part as a "spacer element" and this spacer function plays an important role in directing subgenomic RNA synthesis during virus replication. | Yang D., Liu P., Giedroc D.P., Leibowitz J. Mouse hepatitis virus stem-loop 4 functions as a spacer element required to drive subgenomic RNA synthesis. J. Virol. 2011;85(17):9199–9209. |
| | Stem Loop 4.5 (SL 4.5) | 131-147 | Importantly, the AUG of the predicted upstream ORF (uORF) is phylogenetically conserved among β-coronaviruses and is accessible for recognition by a scanning ribosome. | Stem-loop III in the 5' untranslated region is a cis-acting element in bovine coronavirus defective interfering RNA replication. Raman S, Bouma P, Williams GD, Brian DA. J Virol. 2003 Jun; 77(12):6720-30. |
| 3'-UTR | Bulged Stem Loop (BSL) | Immediately 3' from the N gene stop codon | 68 nts bulged stem-loop downstream of N gene stop codon. SL conserved among the betacoronaviruses and the pairing, but not the primary sequence, is critical. Essential for RNA replication and viral replication. | Hsue B., Masters P.S. A bulged stem-loop structure in the 3' untranslated region of the genome of the coronavirus mouse hepatitis virus is essential for replication. J. Virol. 1997;71(10):7567–7578. |
| | Stem Loop 1 (SL1) | | A hairpin stem-loop which can form a 54 nts hairpin-type pseudoknot, which is required for BCoV DI RNA replication | Williams G.D., Chang R.Y., Brian D.A. A phylogenetically conserved hairpin-type 3' untranslated region pseudoknot functions in coronavirus RNA replication. J. Virol. 1999;73(10):8349–8355. |
| | Poly-A Tail | 3'terminal sequences | Required for negative strand replication; minimum of 5 nts required | Lin Y.J., Liao C.L., Lai M.M. Identification of the cis-acting signal for minus-strand RNA synthesis of a murine coronavirus: implications for the role of minus-strand RNA in RNA replication and transcription. J. Virol. 1994;68(12):8131–8140; Spagnolo J.F., Hogue B.G. (2001) Requirement of the Poly(A) Tail in Coronavirus Genome Replication. In: Lavi E., Weiss S.R., Hingley S.T. (eds) The Nidoviruses. Advances in Experimental Medicine and Biology, vol 494. Springer, Boston, MA. https://doi.org/10.1007/978-1-4615-1325-4_68 |

### Copy-back UTR structures have been seen in defective interfering particles and during synthetic genome assembly

While copy-back structures have never been seen previously in an infective coronavirus, they are found in two settings: during infection, in which defective interfering (DI) particles can be created by errors in RNA translation[13] and during laboratory synthetic genome assembly.

In 5′ copy-back DI genomes a portion of the original virus genome is repeated in a reverse complement form and can be represented as a stem-like structure. This is the BANAL-236 structure. However, defective interfering particles do have an internal packaging signal[14] which is absent in BANAL-236. Even as a defective interfering particle, BANAL-236 has no mechanism of packing the RNA genome into a virion particle. Furthermore, DI particles have not been found that have the full-length genome with copy-back ends like with BANAL-236; usually only a portion of the entire genome is found.

Given the circumstances of BANAL-236, neither of these alternative explanations for the UTR findings are considered appropriate.

**BANAL-236 genome sequence and infectivity has been verified**

The simplest explanation of the findings in Reference 2 would be that either the genome sequence or the viral infectivity experiment was an error. Therefore, the correspondence author of the pre-print was contacted concerning these findings before this manuscript was completed. He confirmed the findings in Reference 2, specifically that BANAL-236 was infective in VERO-E6 cells and that the genome sequence was correct.

Given the potential that BANAL-236 is both the first bat sarbecovirus to directly infect human cells and that it also does so by unique, never before seen methods of replication, translation, and RNA packaging, it is important that a collaborative effort be undertaken to understand the molecular mechanisms it has evolved to allow this to occur. For example, current efforts to create therapeutics against conserved stem-loop structures would be ineffective against BANAL-236.[15]

**Acknowledgements**

I wish to thank Dr. Martin Lee, Dept. of Statistics, UCLA, for the probability determination of the 5'-UTR length of the BANAL-236 genome. I wish to thank Adrian Jones for useful discussions during the preparation of this manuscript.

[13] Beauclair G, Mura M, Combredet C, Tangy F, Jouvenet N, Komarova AV. *DI-tector*: defective interfering viral genomes' detector for next-generation sequencing data. *RNA*. 2018;24(10):1285-1296. doi:10.1261/rna.066910.118

[14] Phylodynamic analysis of the emergence and epidemiological impact of transmissible defective dengue viruses. *Ke R, Aaskov J, Holmes EC, Lloyd-Smith JO PLoS Pathog. 2013 Feb; 9(2):e1003193.*

[15] Lulla V, Wandel MP, Bandyra KJ, Ulferts R, Wu M, Dendooven T, Yang X, Doyle N, Oerum S, Beale R, O'Rourke SM, Randow F, Maier HJ, Scott W, Ding Y, Firth AE, Bloznelyte K, Luisi BF. 2021. Targeting the conserved stem loop 2 motif in the SARS-CoV-2 genome. J Virol 95:e00663-21. https://doi.org/10.1128/JVI .00663-21